\documentstyle[12pt]{article}
\begin{document}
\tolerance=5000
\def\be{\begin{equation}}
\def\ee{\end{equation}}
\def\bea{\begin{eqnarray}}
\def\eea{\end{eqnarray}}
\def\nn{\nonumber \\}
\def\cF{{\cal F}}
\def\det{{\rm det\,}}
\def\Tr{{\rm Tr\,}}
\def\e{{\rm e}}
\def\etal{{\it et al.}}
\def\erp2{{\rm e}^{2\rho}}
\def\erm2{{\rm e}^{-2\rho}}
\def\er4{{\rm e}^{4\rho}}
\def\etal{{\it et al.}}

\  \hfill
\begin{minipage}{3.5cm}
OCHA-PP-163 \\
NDA-FP-81 \\
September 2000 \\
\end{minipage}

\vfill

\begin{center}
{\large\bf Scheme-dependence of holographic conformal anomaly 
in d5 gauged supergravity with non-trivial bulk potential}

\vfill

{\sc Shin'ichi NOJIRI}\footnote{nojiri@cc.nda.ac.jp},
{\sc Sergei D. ODINTSOV}$^{\spadesuit}$\footnote{
On leave from Tomsk Pedagogical University, 
634041 Tomsk, RUSSIA. \\
odintsov@ifug5.ugto.mx, odintsov@mail.tomsknet.ru}, \\
and {\sc Sachiko OGUSHI}$^{\diamondsuit}$\footnote{
JSPS Research Fellow,
g9970503@edu.cc.ocha.ac.jp}
\\

\vfill

{\sl Department of Applied Physics \\
National Defence Academy,
Hashirimizu Yokosuka 239-8686, JAPAN}

\vfill

{\sl $\spadesuit$
Instituto de Fisica de la Universidad de Guanajuato,
Lomas del Bosque 103, Apdo. Postal E-143, 
37150 Leon,Gto., MEXICO }

\vfill

{\sl $\diamondsuit$ Department of Physics,
Ochanomizu University \\
Otsuka, Bunkyou-ku Tokyo 112-0012, JAPAN}

\vfill

{\bf ABSTRACT}

\end{center}

Bosonic sector of d5 gauged supergravity with specific 
parametrization of full scalar coset is considered 
(multi-dilaton gravity). Using holographic RG in the form 
suggested by de Boer-Verlinde-Verlinde the holographic d4 
conformal anomaly (in the sector of curvature invariants) is 
found when bulk potential is not constant. Its comparison 
with the earlier calculation done in the scheme where 
expansion of metric and dilaton over radial 
coordinate of AdS space is used demonstrates scheme dependence of 
holographic conformal anomaly. In AdS/CFT correspondence where 
dilatons play role of coupling constants it coincides with 
multi-loop conformal anomaly which depends on regularization 
scheme in interacting dual QFT. Hence, scheme dependence of 
holographic conformal anomaly is consistent 
with dual QFT expectations.

\newpage

Motivated by AdS/CFT correspondence \cite{AdS} there was 
recently much attention paid to study of RG flows\cite{BC} 
(and refs. therein) from SG side. 
Such investigation gives the possibility to use the solutions of 
multi-dimensional (gauged) supergravity in order to describe RG 
flows in dual CFT living on the boundary of correspondent 
AdS-like spacetime. 
In particulary, classical AdS-like solutions of d5 gauged 
supergravity after the expansion over radial coordinate may 
be also used to get holographic conformal 
anomaly(CA) \cite{HS,SN,anom,LCA} for dual QFT. The calculation of 
holographic CA in such scheme gives very useful check of 
AdS/CFT correspondence (especially, for maximally SUSY 
super Yang-Mills theory). Moreover, as dilaton(s) plays the 
role of gauge coupling constant in AdS/CFT set-up it is 
expected that holographic CA 
(with non-trivial dilaton dependence \cite{SN,LCA,LW}) gives 
useful approximation to exact QFT CA of dual theory. Usually, 
multi-loop quantum calculation is almost impossible to do, the 
result is known only in couple first orders of loop expansion, 
hence use of holographic CA is a challenge. CA for interacting 
QFT may be expressed in terms of gravitational invariants 
multiplied to multi-loop QFT beta-functions (see ref.\cite{ans} 
for recent discussion). One of the features of multi-loop 
beta-functions for coupling constants is their explicit scheme 
dependence (or regularization dependence) which normally 
occurs beyond second loop. This indicates that making 
calculation of holographic CA which corresponds to dual 
interacting QFT in different schemes leads also to scheme 
dependence of such CA. Of course, this should happen in the 
presence of non-trivial dilaton(s) and bulk potential.

Recently, in refs.\cite{DVV} (see also \cite{BGM}) there appeared 
formulation of holographic RG based on Hamilton-Jacobi approach. 
This formalism permits to find the holographic CA without using 
the expansion of metric and dilaton over radial coordinate in 
AdS-like space. The purpose of present Letter is to calculate 
holographic CA for multi-dilaton gravity with non-trivial bulk 
potential in BVV formalism \cite{DVV}. Then, the coefficients 
of curvature invariants as functions of bulk potential are 
obtained. The comparison of these coefficients (c-functions) 
with the ones found earlier\cite{SN,LCA} in the scheme of 
ref.\cite{HS} is done. It shows that coefficients coincide 
only when bulk potential is constant, in other 
words, holographic CA including non-constant bulk potential is scheme 
dependent. 

We start from the 5-dimensional dilatonic gravity action
which is given by
\bea
\label{i}
S={1\over 2\kappa_{5}}\int d^{5}x \sqrt{g}\left[
R+{1\over 2}G(\phi)(\nabla \phi)^{2} +V(\phi) \right].
\eea
and choose the  5-dimensional metric in the following form 
as used in \cite{DVV,LW}
\bea
g_{m n}dx^{m}dx^{n}=d\rho^{2}+\gamma_{\sigma\nu}(\rho,x)
dx^{\sigma}dx^{\nu} 
\eea 
Here $\rho$ is the radial coordinate in AdS-like background.
In the following we only consider the case $G(\phi)=-1$ 
in (\ref{i}) for simplicity. 
As in \cite{DVV,LW}, we adopt Hamilton-Jacobi theory.  
First, we shall cast the 5-dimensional dilatonic gravity action
into the canonical formalism 
\bea
I &=&{1\over 2\kappa_{5}^{2}}\int _{M}d^{5}x\sqrt{g}\left\{
R+{1\over 2}G(\phi)(\nabla \phi)^{2}+V(\phi) \right\}\\
&\equiv& {1\over 2\kappa_{5}^{2}}\int d\rho L \nn
L&=& \int d^{4} x \sqrt{\gamma}
\left[\pi_{\sigma\nu}\dot{\gamma}^{\sigma\nu}
 -\Pi\dot{\phi}-{\cal H} \right]
\eea
where $\dot{}$ denotes the derivative with respect to $\rho$.
The canonical momenta and the Hamiltonian density are defined by
\bea
\label{ham1}
\pi_{\sigma\nu}&\equiv & {1\over \sqrt{\gamma}}
{\delta L\over \delta 
\dot{\gamma}^{\sigma\nu}},\quad \Pi \equiv {1\over \sqrt{\gamma}}
{\delta L\over \delta  \dot{\phi}} \nn
{\cal H}&\equiv& {1\over 3}\pi ^2 -\pi_{\sigma \nu}\pi^{\sigma \nu}
+{\Pi ^2 \over 2G}-{\cal L} ,\\
{\cal L}&\equiv& {\cal R}+{1\over 2}G
\gamma^{\sigma\nu}\partial_{\sigma}\phi
\partial_{\nu}\phi +V .\nonumber
\eea
Here ${\cal R}$ is Ricci scalar of the boundary metric 
$\gamma_{\mu\nu}$. The flow velocity of $g_{\mu \nu} $ is given by
\bea
\dot{\gamma}_{\sigma \nu} = 2\pi _{\sigma \nu}
 -{2\over 3}\gamma_{\sigma \nu}\pi^{\lambda }_{\lambda} .
\eea
This canonical formulation constraints Hamiltonian as 
${\cal H}=0$, which leads to the equation
\bea
\label{hj}
{1\over 3}\pi ^2 -\pi_{\sigma\nu}\pi^{\sigma \nu}+{\Pi ^2 \over 2G}
={\cal R}+{1\over 2}G\gamma^{\sigma \nu}\partial_{\sigma}\phi
\partial_{\nu}\phi +V
\eea 
Applying de Boer-Verlinde-Verlinde formalism, one can decompose
the action $S[\gamma,\phi]$ in a local and non-local part 
as follows 
\bea
\label{act}
S[\gamma,\phi]&=&S_{EH}[\gamma ,\phi]+\Gamma [\gamma,\phi]\nn
S_{EH}[\gamma,\phi]&=&\int d^{4}x\sqrt{\gamma}\left[
Z(\phi)R+{1\over 2}M(\phi)\gamma^{\mu \nu}\partial_{\mu}\phi 
\partial_{\nu}\phi + U(\phi) \right] .
\eea 
Here $S_{EH}$ is tree level renormalized action and $\Gamma$ 
contains the higher-derivative and non-local terms. 
The canonical momenta are related to the Hamilton-Jacobi 
functional $S$ by 
\bea
\label{mom1}
\pi_{\sigma\nu}={1\over \sqrt{\gamma}}
{\delta S \over \delta \gamma^{\sigma \nu}},
\quad \Pi = {1\over \sqrt{\gamma}}
{\delta S \over \delta \phi }
\eea
The expectation value of stress tensor $<T_{\sigma \nu}>$ 
and that of the gauge invariant operator $<O_{\phi}>$ which 
couples to $\phi$ can be related to $\Gamma$ by 
\bea
\label{can}
\left< T_{\sigma \nu} \right>={2 \over \sqrt{\gamma}}
{\delta \Gamma \over \delta \gamma^{\sigma \nu} },
\quad  \left< O_{\phi} \right>= {1\over \sqrt{\gamma}}
{\delta \Gamma \over \delta \phi }\ .
\eea
Then, one can get holographic trace anomaly in the following form
\bea
<T_{\mu}^{\mu}>= \beta \left< O_{\phi} \right>
-c R_{\mu\nu}R^{\mu\nu}+dR^{2}
\eea
where $\beta$ is some beta function and coefficients 
$c$ and $d$ are c-functions. 
Explicit structure of $\beta \left< O_{\phi} \right>$ is given 
in \cite{LCA}:
\bea
\label{bOh}
\beta \left< O_{\phi} \right>
&=&-2\left[h_3 R^{ij}\partial_{i}\phi\partial_{j}\phi 
+ h_4 Rg^{ij}\partial_{i}\phi\partial_{j}\phi
+ h_5 {R \over \sqrt{-g}}\partial_{i}
(\sqrt{-g}g^{ij}\partial_{j}\phi) \right. \nn
&& + h_6 (g^{ij}\partial_{i}\phi\partial_{j}\phi)^2 
+ h_7 \left({1 \over \sqrt{-g}}\partial_{i}
(\sqrt{-g}g^{ij}\partial_{j}\phi)\right)^2 \nn
&& \left. + h_8 g^{kl}\partial_{k}\phi\partial_{l}\phi
{1 \over \sqrt{-g}}\partial_{i}(\sqrt{-g}g^{ij}\partial_{j}\phi)
\right] \ .
\eea
Here $h_3\cdots h_8$ are functions of dilaton $\phi$: 
$h_i=h_i(\phi)$ ($i=3,4,\cdots,8$). 
To get the explicit forms of c-functions, one substitutes 
the action (\ref{act}) into (\ref{mom1})(\ref{can}) thus one can get   
the relation between potentials $U$ and $V$ by using Hamilton-Jacobi
equation (\ref{hj}). From the potential term, we get 
\bea
\label{UV1}
{U^{2} \over 3}+{U'^{2}\over 2G}=V 
\eea
and the curvature term $R$ leads to 
\bea
\label{UV2}
{U\over 3}Z+{U'\over G}Z'=1.
\eea
where $'$ denotes the derivative with respect to $\phi$.  
Examining the terms of $\left<T_{\sigma \nu}\right>$, 
one can get the explicit form of c-functions as
\bea
c={6Z^{2}\over U},\quad d={2\over U}
\left(Z^{2}+{3Z'^{2}\over 2G}\right).
\eea
If we choose constant potential $V(\phi)=12$, 
by using (\ref{UV1}) and (\ref{UV2}), we find  
$U$, $Z$ become constant:
\bea
U=6,\quad Z={1\over 2}
\eea
Then, c-function $c$ and $d$ become 
\be
\label{UVcd}
c={1 \over 4}\ , \quad d={1\over 12} \ .
\ee
This exactly reproduces the correspondent coefficients of 
holographic conformal anomaly obtained in ref.\cite{HS} 
in the scheme where expansion of d5 AdS metric in terms 
of radial AdS coordinate has been adopted.
Now we can understand the coincidence of Weyl anomaly 
calculation between scheme of ref.\cite{HS,SN,LT} and 
de Boer-Verlinde-Verlinde formalism when the scalar potential is 
constant \cite{FMS}.  

At the next step we come to application of above holographic 
RG formalism in the calculation of conformal anomaly for d5 
multi-dilaton gravity with non-trivial bulk potential. 
Such a theory naturally appears as bosonic sector of d5 gauged 
supergravity. We consider maximally SUSY gauged supergravity, 
where scalars parameterize a submanifold of the full scalar coset 
\cite{CGLP,MTR}. 
As a result, the bulk potential cannot be chosen arbitrarily.  
Hence, one limits to the case that includes $N$ scalars and the 
coefficient $G=-1$. 
The bosonic sector of the action in this case is
\bea
\label{mul}
S={1 \over 16\pi G}\int_{M_D} d^Dx \sqrt{-\hat G}
\left\{ \hat R - \sum_{\alpha=1 }^{N} {1 \over 2 }
(\hat\nabla\phi_{\alpha})^2 
+V (\phi_{1},\cdots ,\phi_{N} ) \right\}.&&
\eea
The maximally SUSY supergravity in $D=5$ contains 
$42$ scalars (the construction of such 
d5 gauged supergravity is given in Ref.~\cite{Peter}). 
The maximal supergravity parameterizes the coset $E_{11-D}/K$, 
where $E_{n}$ is the maximally non-compact form of the 
exceptional group $E_{n}$, and $K$ is its maximal 
compact subgroup. The group $SL(N,R)$, a subgroup of $E_{n}$, can 
be parameterized with the coset $SL(N,R)/SO(N)$, and we use the 
local $SO(N)$ transformations in order to diagonalize the 
scalar potential $V(\phi )$ as in ref.\cite{CGLP} 
\bea
\label{CGL}
V={(D-1)(D-2) \over N(N-2)}\left(\left(
\sum_{i=1}^{N}X_{i}\right)^{2}
 -2\left(\sum_{i=1}^{N}X_{i}^{2} \right) \right) .
\eea
Especially for $D=5$, we have
\bea
V={1\over 2}\left\{ \left( \sum_{i=1}^{6} X_{i}\right)^{2}
-2 \left( \sum_{i=1}^{6}X_{i}^{2} \right) \right\}
\eea
Let us briefly describe the parameterization leading to the 
action of form (\ref{mul}) given in Ref.~\cite{CGLP}. 
In the above gauged supergravity case, in $D=5$ one should 
set $N=6$. The $N$ scalars $X_{i}$, which are constrained by 
\bea
\prod_{i=1}^{N}X_{i}=1\ ,
\eea
can be parameterized in terms of $(N-1)$ independent
dilatonic scalars $\phi_{\alpha}$ as 
\bea
\label{bb1}
X_{i}=\e^{-{1\over 2}b^{\alpha}_{i}\phi_{\alpha}}
\eea
Here the quantities $b_{i}^{\alpha}$ are the weight vectors of 
the fundamental representation of $SL(N,R)$, which satisfy
\bea
\label{bb2}
&& b_{i}^{\alpha}b_{j}^{\alpha}=8 \delta_{ij} -{8 \over N},\quad 
\sum_{i}b_{i}^{\alpha} =0 \nn
&& b_{i}^{\alpha}b_{i}^{\beta}=4(N-4)\delta^{\alpha \beta}\ .
\eea
Then, potential has a minimum at $X_{i}=1$ ($N>5$) at the point 
$\phi_{\alpha}=0$, where $V=(D-1)(D-2)$. 

To get c-functions $c$ and $d$, one take $U$
and $Z$ as
\bea
\label{UZK}
U &=& A\sum_{i=1}^{6}\e^{-{1\over 2}b_{i}^{\alpha}\phi_{\alpha}}=
A\sum_{i=1}^{6} X_{i}\nn
Z &=& B\sum_{i=1}^{6}\e^{{1\over 2}b_{i}^{\alpha}\phi_{\alpha}}=
B\sum_{i=1}^{6} X_{i}^{-1}
\eea
where $A$ and $B$ can be determined by the analogues 
with $G=-1$ of the conditions (\ref{UV1}) and 
(\ref{UV2}) \cite{FMS}, 
\be
\label{UVm}
{U^{2} \over 3} - \sum_\alpha \left({\partial U 
\over \partial \phi_\alpha} \right)^{2}=V \ ,
\quad {U\over 3}Z - \sum_\alpha 
{\partial U \over \partial \phi_\alpha}
{\partial Z \over \partial\phi_\alpha} =1.
\ee
as follows 
\bea
A=\pm 1,\quad B=\pm {1\over 12} .
\eea
Then, using (\ref{UVcd}), c-functions are found
\bea
\label{cd}
c&=&{6Z^{2}\over U}={1\over 24}\left(
\sum_{i}^{6} X_{i}^{-1} \right)^{2}
\left( \sum_{j}^{6} X_{j} \right)^{-1} \\
d&=&{2\over U}\left(Z^{2} - {3 \over 2}\sum_\alpha 
\left({\partial Z \over \partial \phi_\alpha}\right)^{2}\right) \nn
&=&{1\over 24}\left( \sum_{j}^{6} X_{j} \right)^{-1} 
\left\{ {1\over 2}
\left(\sum_{i}^{6}  X_{i}^{-1} \right)^{2}
 -\sum_{i}^{6}  X_{i}^{-2} \right\}
\nonumber
\eea
Thus, the coefficients of gravitational terms in
holographic conformal anomaly (corresponding to dual CFT) 
from multi-dilaton five-dimensional gravity are found. The 
holographic RG formalism is used in such calculation.
In \cite{LW}, the c-functions $c$ and $d$ have been found for 
version of dilaton gravity dual to non-commutative Yang-MIlls (NCYM) 
theory. The gravity theory contains only one dilaton field 
$\phi$. The action can be obtained by putting 
\be
\label{NCYM}
G(\phi)=-{20 \over 3\phi^2}\ ,\quad
V(\phi)={1 \over \phi^{8 \over 3}}\left(20 - {8 \over 
\phi^4}\right). 
\ee
Then  using (\ref{UV1}), (\ref{UV2}) and (\ref{UVcd}), 
one finds
\be
\label{NCYM2}
c={\phi^2(\phi^2 + 2 )^2 \over 12(5\phi^2 -2)}\ ,\quad
d={\phi^2(\phi^4 + 8\phi^2 + 6) \over 60(5\phi^2 -2)}\ .
\ee
This coincides with result of ref.\cite{LW} and gives 
useful check of our calculation.

Let us turn now to results of calculation of holographic 
conformal anomaly done in refs.\cite{SN,LCA} where another 
scheme \cite{HS} was used. In such scheme five-dimensional 
metric and scalars are expanded in terms of radial fifth 
coordinate (for more detail, see \cite{LCA}). Note that 
as in above evaluation the dilaton and bulk potential are 
considered to be non-trivial and non-constant.

The functions $2h_{1}$ and $-2h_{2}$ in notations of 
ref.\cite{LCA} correspond to c-functions $d$ and $c$ in 
(\ref{cd}), respectively, and they are given by, 
\bea
\label{hh}
h_{1}&=&{3(62208+22464{\cal V}+2196{\cal V}^{2}
+72{\cal V}^{3}+{\cal V}^{4})l^{3}\over
16(6+{\cal V})^{2}(18+{\cal V})^{2}(24+{\cal V})} \\
h_{2}&=&-{3 (288+72{\cal V}+{\cal V}^{2})l^{3}\over 
8(6+{\cal V})^{2}(24+{\cal V})}
\eea
where
\bea
\label{calv}
{\cal V}\equiv V(\phi)-V(0)=V(\phi)-12.
\eea

Since the expressions of $c$, $d$ seem to be very different 
from $2h_1$, $-2h_2$ which are obtained with help of expansion 
of metric and bulk potential on radial coordinate, we 
now investigate if they are really different by expanding 
$X_{i}$ on $\phi$ (up to second order on $\phi ^{2}$)
\bea
X_{i}&=&1-{1\over 2}b_{i}^{\alpha}\phi_{\alpha}
+{1\over 8}(b_{i}^{\alpha}\phi_{\alpha})^{2}\nn
\sum_{i}^{6} X_{i}&=& 6-{1\over 2}\sum_{i}^{6}b_{i}^{\alpha}
\phi_{\alpha}+{1\over 8}\sum_{i}^{6}(b_{i}^{\alpha}\phi_{\alpha})^{2}\nn
(\sum_{i}^{6} X_{i})^{2}&=& 36-6\sum_{i}^{6}b_{i}^{\alpha}
\phi_{\alpha}+{3\over 2}\sum_{i}^{6}(b_{i}^{\alpha}\phi_{\alpha})^{2}
+{1\over 4}(\sum_{i}^{6}b_{i}^{\alpha}
\phi_{\alpha})^{2}\nn
\sum_{i}^{6} X_{i}^{2} &=& 6-\sum_{i}^{6}b_{i}^{\alpha}
\phi_{\alpha}+{1\over 2}\sum_{i}^{6}(b_{i}^{\alpha}\phi_{\alpha})^{2}\nn
X_{i}^{-1}&=&1+{1\over 2}b_{i}^{\alpha}\phi_{\alpha}
-{1\over 8}(b_{i}^{\alpha}\phi_{\alpha})^{2}\nn
\sum_{i}^{6} X_{i}^{-1}&=& 6+{1\over 2}\sum_{i}^{6}b_{i}^{\alpha}
\phi_{\alpha}-{1\over 8}\sum_{i}^{6}(b_{i}^{\alpha}\phi_{\alpha})^{2}\nn
(\sum_{i}^{6} X_{i}^{-1})^{2}&=& 36+6\sum_{i}^{6}b_{i}^{\alpha}
\phi_{\alpha}-{3\over 2}\sum_{i}^{6}(b_{i}^{\alpha}\phi_{\alpha})^{2}
+{1\over 4}(\sum_{i}^{6}b_{i}^{\alpha}
\phi_{\alpha})^{2}
\eea
Using (\ref{bb1}) and (\ref{bb2}), one finds c-functions $c$ and $d$ 
in (\ref{cd}) are given by
\bea
\label{cfun}
c&=&{1\over 4} 
-{1\over 64}\sum_{i}^{6}(b_{i}^{\alpha}\phi_{\alpha})^{2} \\
d&=&{1\over 12}-{1\over 144}
\sum_{i}^{6}(b_{i}^{\alpha}\phi_{\alpha})^{2}
\eea
$h_{1}$ and $h_{2}$ in (\ref{hh}) are given by
\bea
h_{1}&=&{1\over 2}\left( {1\over 12}
 -{1\over 384}\sum_{i}^{6}(b_{i}^{\alpha}
\phi_{\alpha})^{2} \right)\\
-h_{2}&=&{1\over 2}\left( {1\over 4} -{1\over 128}\sum_{i}^{6}(b_{i}^{\alpha}
\phi_{\alpha})^{2} \right)\ .
\eea
Then $2h_1$ and $-2h_2$ do not coincide with $d$ and $c$, except 
the leading constant part. One finds the formalism by 
de Boer-Verlinde-Verlinde \cite{DVV} does not reproduce the result 
based on the scheme of ref.\cite{HS}. 
Technically, this disagreement might occur since we expand dilatonic 
potential in the power series on $\rho$ and this is the reason 
of ambiguity and scheme dependence of holographic conformal anomaly.
 
This result means that holographic CA (with non-trivial bulk 
potential and non-constant dilaton) is scheme dependent. 
AdS/CFT correspondence says that such holographic CA 
should correspond to (multi-loop) QFT CA (dilatons play the role of 
coupling constants). However, QFT multi-loop CA depends on 
regularization (as beta-functions are also scheme dependent). 
Hence, scheme dependence of holographic CA is consistent with 
QFT expectations. That also means that two different formalisms 
we discussed in this Letter actually should correspond to different 
regularizations of dual QFT.\\

\noindent{\bf Acknowledgements}

The work of S.O. has been supported in part by Japan Society 
for the Promotion of Science and that of S.D.O. by CONACyT (CP, 
ref.990356 and grant 28454E).  

\newpage

\appendix


\noindent
{\bf Appendix} 

\ 

In this Appendix,
in order to show that scheme dependence takes place in other dimensions we 
consider calculation of 2d holographic conformal anomaly from 
3d dilatonic gravity with arbitrary bulk potential.
Using Hamilton-Jacobi formalism similarly to 5-dimensional 
gravity, we cast 3d ADM Hamiltonian
density as (instead of (\ref{ham1})) 
\be
\label{ham2}
{\cal H} \equiv  \pi ^2 -\pi_{\sigma \nu}\pi^{\sigma \nu}
+{\Pi ^2 \over 2G}-{\cal L} \; .
\ee
Hamiltonian constraint: ${\cal H}=0$, leads to the following
equation  similar to (\ref{hj}) 
\be
\label{hj2}
\pi ^2 -\pi_{\sigma\nu}\pi^{\sigma \nu}+{\Pi ^2 \over 2G}
={\cal R}+{1\over 2}G\gamma^{\sigma \nu}\partial_{\sigma}\phi
\partial_{\nu}\phi +V
\ee 
We assume the form of 2-dimensional Weyl anomaly as
\be
<T_{\mu}^{\mu}>= \beta \left< O_{\phi} \right>+c R \; ,
\ee 
where $R_{\mu \nu}={1\over 2}g_{\mu\nu}R$.  
To solve the Hamilton-Jacobi equation (\ref{hj2}), one uses
 the same procedure  as in 4d.
Substituting Hamilton momenta (\ref{mom1}), (\ref{can}) 
into (\ref{hj2}), we obtain the relation between $U$ and $V$
from the potential term 
\be
\label{UV3}
{U^{2} \over 2}+{U'^{2}\over 2G}=V , 
\ee
and the curvature term $R$ leads to the central charge $c$ 
\be
\label{cent}
c=-{2\over U}\left( 1-{Z'U' \over G}\right) .
\ee
We also obtain the following equation from $R^2$ term:
\be
\label{ZZ}
{c^2 \over 4}+{{Z'}^2 \over G}=0\ .
\ee
By deleting $Z'$ from (\ref{cent}) by using (\ref{ZZ}), we 
find the following expressin for the c-function $c$
\be
\label{cc}
c=-{2 \over U \pm {U' \over \sqrt{-G}}}\ .
\ee
Especially if we choose constant potential $V(\phi)=2$ and 
$G=-1$, we find
$U$ and $c$ from eqs.(\ref{UV3}) and (\ref{cc}),
\be
U= \pm 2,\quad c =\mp 1 \; .  
\ee   
Taking $c=1$, this holographic RG result (at constant bulk 
potential) exactly agrees with the one of previous 
work \cite{LCA} (where expansion around of AdS space 
was used).

Next we consider 3-dimensional bulk potential as 
\be
\label{pot}
V={2\over \cosh^2\phi}\ ,
\ee
with $G=-1$. 
Then by using (\ref{UV3}) and (\ref{cc}), we obtain
\be
\label{CC1}
U=-{2 \over \cosh\phi}\ ,\quad c=\e^{\pm \phi}\cosh^2\phi\ .
\ee
In \cite{LCA} which is based on expansion method we got 
c-function as 
\bea
\label{CC2}
c_{\rm NOO}&=& \left({{\cal V}(\phi)\over 2}
+2 \right) \left({\cal V}(\phi)+2 \right)^{-1}\\
{\cal V}(\phi) &=& V(\phi)-2 . \nonumber
\eea 
Substituting the potential (\ref{pot}) into (\ref{CC2}),
we obtained c-function as follows:
\be
\label{CC3}
c_{\rm NOO}= {1 + \cosh^2\phi \over 2}\ .
\ee
This c-function (\ref{CC3}) does not coincide with (\ref{CC1})
like in 4d case (apart from the leading, constant part). 
In (\ref{hh}) for 4d case and (\ref{CC2}) for 2d case, the terms 
containing the derivatives of the potential $V$ with respect to 
the scalar field $\phi$ were neglected \cite{LCA}. As one might 
doubt that this might be the origin of the above disagreement, 
we investigate 2d case explicitly. If we include the neglected 
terms, $c_{\rm NOO}$ in (\ref{CC2}) is modified as follows
\bea
\label{CC4}
\tilde c_{\rm NOO}&=& 1 + {{2{\cal V}'(\phi) \over 
{\cal V}''(\phi)\left({\cal V}(\phi) + 2\right)
 - \left({\cal V}'(\phi)\right)^2} - {\cal V}(\phi)
\over 2\left({\cal V}(\phi) +2\right) } \ .
\eea 
By substituting the potential (\ref{pot}) into (\ref{CC4}),
we obtained modified c-function $\tilde c_{\rm NOO}$ as follows:
\be
\label{CC5}
\tilde c_{\rm NOO}= {1 + \cosh^2\phi \over 2}
+ {1 \over 4}\sinh\phi \cosh^5\phi\ .
\ee
This c-function (\ref{CC5}) does not coincide with (\ref{CC1})
again.

\end{document}